\documentclass[12pt]{iopart}
\eqnobysec
\usepackage{graphicx} 
\usepackage{color}
\usepackage{multirow}

\begin{document}
\title[Thermostatistics of small systems]{Thermostatistics of small systems: Exact results in the microcanonical formalism}

\author{E N Miranda$^{1,2,4}$, Dal\'{i}a S Bertoldi$^{1,3}$}

\address{$^1$ Instituto de Ciencias B\'{a}sicas, Universidad Nacional de Cuyo 5500 - Mendoza, Argentina}
\address{$^3$ IANIGLA - CONICET, CCT Mendoza, 5500 - Mendoza, Argentina}
\address{$^2$ CONICET - Mendoza, Argentina}
\address{$^4$ Departamento de F\'{i}sica, Universidad Nacional de San Luis 5700- San Luis, Argentina}
\ead{emiranda@mendoza-conicet.gov.ar}

\begin{abstract}
Several approximations are made to study the microcanonical formalism that are valid in the thermodynamics limit. Usually it is assumed that: 1) Stirling's approximation can be used to evaluate the number of microstates; 2) the surface entropy can be replaced by the volume entropy; 3) derivatives can be used even if the energy is not a continuous variable. And it is also assumed that the results obtained in the microcanonical formalism agree with those from the canonical one. However it is not clear if these assumptions are right for very small systems (10--100 particles). To answer this question, two systems with exact solutions (the Einstein model and the two-level system) have been solved with and without these approximations.
\end{abstract}
\pacs{1.40, 5.20, 5.40}
\submitto{\EJP}
\maketitle

\section{Introduction}
Rapid experimental and computational developments in the last decades have generated an increasing interest in the application of notions from Thermodynamics and Statistical Mechanics to small systems\cite{Adib,Hill}, even those on the order of $10\leq N \simeq 100$, where \textit{N} is the number of particles.  The microcanonical formalism allows the application of Statistical Mechanics to isolated small systems, but there are fundamental principles of Thermostatistics that have to be reconsidered\cite{Gross,Casetti}.

The interest in small systems is not limited to Physics, but also extends to Biology and Chemistry, going from clusters\cite{Baletto,Berry}, to thin films\cite{Ohring} and magnetic nanoparticles\cite{Lu,Young}, and up to biological molecules of a few nanometers\cite{Itamar}. These systems have properties different from those of macroscopic size. The phase transitions\cite{Gross, Dunkel,Berry}, the thermal conductivity\cite{Chantrenne, Liang}, the fluctuations of the thermodynamic variables\cite{Hill2}, the specific heat \cite{Carignano, Behringer}, etc., are all affected by the small size of the system.

In standard Statistical Mechanics, several approximations are made but their validity is not clear for small systems. Among these approximations, one should note the use of Stirling's formula and the replacement of the surface entropy by the volume one. It is also assumed that the results in the microcanonical and canonical formalism are equivalent. However for systems with few particles, different statistical ensembles give different results\cite{Kastner,Touchette,Dunkel}. In the case of ensemble non-equivalence, instead of choosing the statistical ensemble according to convenience, this choice has to reflect the physical situation of interest, e.g., an energetically isolated cluster should be studied within the microcanonical formalism. Hence it is important to determine from which size the properties of the small system match those found in the thermodynamic limit. 

In this paper, these questions are answered for two known models that admit exact analytical solutions: the two-level system and the quantum oscillator or Einstein solid. These models are typical examples of a Statistical Mechanics course, so that the present article has a clear teaching interest: fundamental questions of  Statistical Physics are analyzed with tools available to advanced undergraduate students. A first step in this direction can be found in Ref. 19.

The structure of this article is the following: In Section 2, the main questions to be answered are stated. In Sections 3 and 4, the two-level system and the Einstein solid are studied in detail. Finally, in Section 5 the obtained results are summarized.

\section{The questions to be addressed}

\subsection{Surface and volume entropy}

If \textit{S} is the entropy, $k_{B}$ the Boltzmann constant, and $\Omega $ the number of microstates accessible to the system, the fundamental equation of statistical mechanics is

\begin{equation}
S=k_{B}\,ln\,\Omega . \label{eq.entropy.omega}
\end{equation}

From Eq.(\ref{eq.entropy.omega}), the microcanonical formalism of Statistical Mechanics is developed. It is not an easy task to evaluate $\Omega $: that evaluation can be carried out for only a few cases. It should be remarked that $\Omega$ is the number of quantum states with energy \textit{E} and $E+\Delta E$ where \textit{E} is a discrete quantity, i.e., it is not a real-valued variable. The evaluation  of $\Omega $ is the enumeration of quantum states inside a very thin spherical shell with radius $E$. Using Eq.(\ref{eq.entropy.omega}), one gets the so-called surface (superscript \textit{sur}) entropy ($S^{sur}$) to use the accepted terminology \cite{Campisi, Adib}. However, for systems in the thermodynamic limit, the number of microstates is so large that the energy can be considered as a continuous quantity; moreover the volume of the shell and that of the hypersphere are almost equal \cite{Callen}. For these reasons, counting microstates is equivalent to evaluating a spherical volume in a high dimensional space. In this way, one gets the volume (superscript \textit{vol}) entropy ($S^{vol}$). In mathematical terms, these entropies are defined as follows:

\begin{equation}
S^{sur}(N,E)=k_{B}ln\int \Omega (E^{\prime },N)\delta (E-E^{\prime})dE^{\prime } ,  
\label{eq.entropy.sur}
\end{equation}

\begin{equation}
S^{vol}(N,E)=k_{B}ln\int \Omega (E^{\prime },N)\theta(E-E^{\prime})dE^{\prime } , 
\label{eq.entropy.vol}
\end{equation}
where $\theta (x)$ is the Heaviside step function and $\delta (x)$ is the Dirac delta function.

In the thermodynamic limit, $S^{vol}=S^{sur}$. However, one may wonder if this also holds for small systems ($10\leq N\simeq 100$). Therefore the first question to be addressed is:
\\
\textbf{Question 1}: If $N\leq 100$, is $S^{sur}(N,E)=S^{vol}(N,E)$?

\subsection{Approximate and exact entropy}

To solve many examples in the microcanonical formalism, some combinatorics should be used and factorials appear naturally. The standard procedure is to apply Stirling's approximation to deal with them: $\ln X!\simeq X\,\ln X-X$. In the thermodynamic limit this is a good trick but it is not clear if it works for small systems. In a previous paper \cite{Bertoldi}, it has been shown that it is inaccurate for systems with $N\sim 100$. In this article, that analysis is refined. In \cite{Bertoldi}, the gamma function was used to replace the logarithm of the factorial and the digamma function was applied for the derivatives. In this paper, the factorial itself is used and the derivatives are replaced by finite differences. Let us call $S_{m,ex} $ the exact (subscrip \textit{ex}) entropy and $S_{m,app}$ the approximate (subscript \textit{app}) one, i.e., using Stirling's formula, both in the microcanonical (subscript \textit{m}) formalism. The second question is:
\\
\textbf{Question 2}: If $N\leq 100$, is $S_{m,ex}(N,E)=S_{m,app}(N,E)$?

\subsection{Canonical and microcanonical temperature and specific heat}

The temperature can be calculated from the entropy, using a well known thermodynamical relation:

\begin{equation}
\frac{1}{T}=\frac{\partial S}{\partial E} .  
\label{eq.temperature}
\end{equation}

It is well known that energy is a discrete quantity, but in the standard calculations it is treated as a continuous variable and its derivatives are calculated in the usual way. This is a reasonable assumption for large systems that have a huge number of energy quanta (\textit{M}) but it may be inexact for systems with low energy and small size. The number of quanta is \textit{M} or \textit{M-1}; for this reason in small systems, derivatives have to be replaced by finite differences and Eq.~(\ref{eq.temperature}) becomes

\begin{equation}
\frac{1}{T(N,M)}=\frac{S(N,M)-S(N,M-1)}{M-(M-1)}=S(N,M)-S(N,M-1).
\label{eq.temperature.finite}
\end{equation}

Equation (\ref{eq.temperature.finite}) is exact since it reflects the discrete nature of energy. Notice that the energy can always be written as $E=M$ by choosing appropriate units. It can be argued that changing a derivative to a finite difference can be done in different ways and Eq. (\ref{eq.temperature.finite}) could also be written as a forward difference $1/T(N,M)=S(N,M+1)-S(N,M)$. However this equation leads to the wrong result since the temperature is different from zero even if the energy is zero: $1/T(N,0)=S(N,1)$. If a backward difference is used, $1/T(N,0)=-S(N,-1)\rightarrow \infty $ because $\Omega (N,-1)$, i.e., the number of microstates, is strictly zero. For this reason, a backward difference scheme is the right choice.

The same considerations are valid for the specific heat. The usual expression is

\begin{equation}
C=\frac{\partial E}{\partial T}. 
\label{eq.Cv}
\end{equation}
Due to the discrete nature of energy, this formula should be replaced by

\begin{equation}
C(N,M)=\frac{M-(M-1)}{T(N,M)-T(N,M-1)}=\left[ T(N,M)-T(N,M-1)\right] ^{-1}.
\label{eq.Cv.finite}
\end{equation}
A backward difference scheme is adopted for the specific heat for the same reason as in the case of temperature.

Let us write $T_{m,app}$ for the temperature obtained using $S_{m,app}$ and Eq. (\ref{eq.temperature}), and $T_{m,ex}$ for the temperature one gets using $S_{m,ex}$ and Eq. (\ref{eq.temperature.finite}). This means that $T_{m,ex}$ has been obtained without Stirling's approximation and takes into account that the number \textit{M} of energy quanta is discrete. In the same way, the specific heat obtained using $S_{m,app}$ and Eq. (\ref{eq.Cv}) is written as $C_{m,app}$, while $C_{m,ex}$ is the one from Eq. (\ref{eq.Cv.finite}) with $S_{m,ex}$. It is a textbook exercise\cite{Callen} to show that the temperature and specific heat obtained in the microcanonical formalism with the usual approximations, which are called $T_{m,app}$ and  $C_{m,app}$ in this article, are the same as those obtained with the canonical formalism (suscript $can$):  $T_{m,app}=T_{can}$ and $C_{m,app}=C_{can}$ for large systems. The final question is related to the equivalence between the results of the microcanonical and canonical formalism.  The question is:
\\
\textbf{Question 3}: If $N\leq 100$, is $T_{m,ex}=T_{can}$ and $C_{m,ex}=C_{can}$?

The aim of the following two sections is to answer the posed questions for the two-level system and the Einstein model (or quantum oscillator).

\section{Two-level systems}    

There are \textit{N} particles, each one with two energy levels: the ground state with energy zero and an excited one with energy $\varepsilon $. The energy of the whole system is $M\varepsilon $, where \textit{M} is the number of energy quanta. To simplify the calculations, it is assumed that $\varepsilon =1$; in this way the system energy is just \textit{M} (yo borraria lo que esta despues del punto y coma). To evaluate the number of microstates one should calculate the number of ways \textit{M} quanta can be distributed among \textit{N} systems: $N!/\left[ M!(N-M)!\right]$.

\subsection{Entropy}

For the sake of simplicity, we put $k_{B}=1$ from now on. According to (\ref{eq.entropy.omega}) and (\ref{eq.entropy.sur}), the entropy becomes 

\begin{equation}
S_{m,ex}^{sur}(N,M)=ln\left[ \frac{N!}{M!(N-M)!}\right].
\label{entropy.ex.sur.twolevel}
\end{equation}

Equation (\ref{entropy.ex.sur.twolevel}) is the exact entropy evaluated in the microcanonical formalism and it is identified with the surface entropy because the total energy of any microstate is exactly \textit{M}. The exact volume entropy is evaluated by adding all the microstates with energy less than or equal to \textit{M}.
(Aca estamos siendo poco claros con la notacion: M es el numero de cuantos de energia no el cuanto con energia maxima. La entropia superficial se calcula con el mayor valor de M mientras que la volumetrica con todos los valores de M. Tal vez podamos llamar $M_max$ al que se usa en la entropia superficial?.) 

\begin{equation}
S_{m,ex}^{vol}=ln\left[ \sum_{m=0}^{M}\frac{N!}{m!(N-m)!}\right].
\label{entropy.ex.vol.twolevel}
\end{equation}

The approximate entropy is obtained by using Stirling's formula to replace the logarithm of the factorials:

\begin{equation}
S_{m,app}=NlnN-MlnM-(N-M)ln(N-M)  .
\label{entropy.app.twolevel}
\end{equation}

It should be noticed that the approximate entropy is equivalent to the canonical one $S_{can}$. It is a textbook exercise \cite{Callen} to show that if  one starts with the partition function, obtains the Helmholtz energy, calculates the derivative with respect to the temperature, one then gets $S_{can}$ in terms of \textit{T}. This temperature has to be rewritten in terms of the natural variables of the microcanonical formalism, i.e., \textit{N} and \textit{M}, as explained below. It comes out that, for two-level systems, it is $S_{can}=S_{m,app}$.

In Figure 1, Eqs. (\ref{entropy.ex.sur.twolevel}), (\ref{entropy.ex.vol.twolevel}), and (\ref{entropy.app.twolevel}) have been plotted for (a) $N=10$ and (b) $N=100$. For a very small system ($N=10$), there is a noticeable difference between the three entropies. Obviously the surface entropy is lower than the volume one because the latter counts more microstates. The canonical (or approximate microcanonical) entropy is the larger one. Note that the difference between the entropies increases with increasing energy. For larger systems ($N \geq 100$), the difference between the three entropies is much smaller. Therefore it is valid to say that the surface entropy can be replaced by the volume one for a system with a few hundred elements. From now on, $S_{m,ex}$ designates the surface entropy evaluated exactly in the microcanonical formalism.

A relative error is introduced to quantify the differences between the canonical and microcanonical results and it is defined as: \textit{(canonical magnitude - microcanonical magnitude)*100 \nolinebreak / microcanonical magnitude.} 

In Table 1, the relative error of the entropy is shown for different system sizes and energies. There is a significant difference ($>25\%$)  for very small systems, but this diminishes with increasing \textit{N}. It cannot be said that the canonical and microcanonical approaches lead to the same results for $N \leq 200$. For larger \textit{N}, the difference becomes negligible.

\subsection{Temperature}

To see whether the microcanonical temperature $T_{m,ex}$ agrees with the canonical one $T_{can}$, the magnitudes in the canonical formalism should be expressed in terms of the energy that is a natural variable of the microcanonical ensemble.

The average energy of a two-level system evaluated within the framework of the canonical formalism is \cite{Kubo}

\begin{equation}
\bar{E}=\left( 1+e^{1/T}\right) ^{-1}.
\end{equation}

In the microcanonical formalism, the average energy is $M/N$. Taking this into account, from the above equation one gets

\begin{equation}
T_{can}=\frac{1}{ln\left[ N/M-1\right] }  .
\label{eq.T.can}
\end{equation}

The exact temperature evaluated in the microcanonical formalism is obtained from (\ref{eq.temperature.finite}) and (\ref{entropy.ex.sur.twolevel}):

\begin{eqnarray}
T_{m,ex}& =\left[ ln\left( \frac{N!}{M!(N-M)!}\right) -ln\left( \frac{N!}{(M-1)!(N-M+1)!}\right) \right] ^{-1} \nonumber \\
&=\frac{1}{ln\left[ N/M-1+1/M\right] } . 
\label{eq.T.ex}
\end{eqnarray}

From (\ref{eq.T.can}) and (\ref{eq.T.ex}), it is clear that the exact microcanonical temperature differs  from the canonical one by a term in \textit{1/M}, i.e., for high energies, both formalisms are equivalent.  If \textit{N} or \textit{M} is small, the two temperatures are different.

In Figure 2, $T_{m,ex}$ and $T_{can}$ are plotted for (a) $N=10$ and (b) $N=100$. In the first case, the difference is relevant for high energies ($M/N>0.2$) but is not noticeable for the larger system ($N=100$) if $M/N<0.5$. The canonical temperature is not defined for $M/N=0.5$ (actually it diverges) but the microcanonical one is a well behaved magnitude for this energy. The meaning of `canonical temperature' in this paper has to be made precise. It means expressing, in terms of \textit{N} and \textit{E} (natural variables of the micocanonical formalism), the temperature one gets in the framework of the canonical formalism.  However, the system is not in contact with a heat reservoir at that temperature.

The exact microcanonical temperature is always well defined for $M/N=0.5$: for $N=10$ it is $T_{m,ex}=5.48$ while for $N=100 $ it is $T_{m,ex}=54.5$. This is an example of the well known fact that no magnitude diverges in a finite size system. 

In Table I, the relative error of the temperature is shown for different system sizes and average energies. For energies greater than $0.2 $ and small system size, the relative errors between the canonical and microcanonical results are relevant ($\> 8\%$): for $M/N=0.4$ and $N=10$ the discrepancy is around $38\%$. Even for $N=200$ and $M/N=0.4$ there is a discrepancy of $2\%$. It is clear that for very small systems, the canonical and microcanonical formalisms lead to different results over the entire energy range.

\subsection{Specific heat}

The specific heat evaluated in the canonical formalism $C_{can}$ is \cite{Callen,Kubo,Reif}

\begin{equation}
C_{can}=\frac{N}{T^{2}}\frac{e^{1/T}}{\left( 1+e^{1/T}\right) },
\label{eq.Cv.can}
\end{equation}

This expression has to be rewritten in terms of \textit{N} and \textit{M}. From (\ref{eq.T.can}) and (\ref{eq.Cv.can}) one gets

\begin{equation}
C_{can}=\frac{M^{2}}{N}\left( \frac{N}{M}-1\right) \left( ln\left[ \frac{N}{M}-1\right] \right) ^{2}  .
\label{eq.Cv.can.micro}
\end{equation}

The exact specific heat evaluated in the microcanonical framework $C_{m,ex}$ is obtained from (\ref{eq.Cv.finite}) using the temperature given by (\ref{eq.T.ex}). In Figure 3, $C_{can}$ and $C_{m,ex}$ are shown for (a) $N=10$ and (b) $N=100$.  If the system is very small ($N=10$), there is an appreciable difference between the two results and a shift in the Schottky bump. For a larger system ($N=100$), both approaches agree except for very low energies ($M/N<0.05$).

In Table 1, the relative error of the specific heat is shown for different system sizes and energies. For $N=10$, the differences are relevant over the entire energy range: for $M/N=0.4$, the error is larger than $45\%$. Again, for such a system it is not true that the canonical and microcanonical formalisms lead to the same results. However, the disparity diminishes with the system size, and for $N=200$, the differences are less than $1\%$.

In Figure 4, the relative errors of the temperature and the specific heat are plotted in terms of \textit{N} for $N/M=0.4$. For $N \leq 40$, the error is larger than $5\%$ for both magnitudes.

\subsection{Summary}

There are some points from among the above results worth stressing. 

\begin{enumerate}
\item  For small systems ($N=10$), the exact surface entropy, the exact volume entropy, and the approximate entropy differ significantly. The relative error between the exact and approximate entropy is always greater than $25\%$; it increases with diminishing energy ($40\%$ for $M/N = 0.1$). The difference between the volumetric and exact entropies behaves in the opposite way: it increases with increasing energy ($4\%$ for $M/N=0.1$ and $17\%$ for $M/N=0.5$). It should be remembered that the approximate entropy is the same as the canonical entropy.
\item For $N=10$, the temperature and the specific heat evaluated exactly in the microcanonical formalism do not coincide with those evaluated in the canonical formalism.
\item  There is a qualitative difference in the temperature at $M/N=0.5$: the canonical temperature diverges while the microcanonical one reaches a finite maximum.
\item There is a shift in the position of the Schottky bump that is noticeable for $N=10$. The maximum is shifted one unit, i.e., one energy quantum, when the specific heat is evaluated in an exact way. 
\item For $N\geq 200$, these differences diminish and can be neglected.
\end{enumerate}

\section{Einstein model of a solid}
The Einstein model is a set of \textit{N} atoms, each one associated with three quantum oscillators with frequency $\omega $. Therefore, there are \textit{3N} oscillators and the total energy is $M\hbar\omega $. Put $\hbar \omega =1$ for simplicity. Now the energy is simply \textit{M}, i.e., the number of quanta. The number of microstates can be obtained from via some combinatorics \cite{Callen, Kubo, Reif}: \textit{M} balls have to be distributed among \textit{3N} boxes. The boxes can be represented as vertical lines and the problem reduces to that of counting the different combinations of $3N+M-1$ elements. It should be noticed that the first element has to be a vertical line and the number of elements to be permuted is reduced by one.

\subsection{Entropy}

From the above considerations, the exact entropy is

\begin{equation}
S_{m,ex}^{sur}(N,M)=ln\left[ \frac{(3N-1+M)!}{(3N-1)!M!}\right].
\label{eq.S.ex.Solid}
\end{equation}

In this expression, only the states with energy \textit{M} are counted, therefore according to Eq. (\ref{eq.entropy.sur}) it is the surface entropy. The volume entropy results from counting all the microstates with $E\leq M$:

\begin{equation}
S_{m,ex}^{vol}(N,M)=ln\left[ \sum_{m=0}^{M}\left( \frac{(3N-1+m)!}{(3N-1)!m!}\right) \right]
\end{equation}

The usual way to get the approximate entropy (found in textbooks) is to apply Stirling's formula ($\ln x!=x\ln x-x$) to Eq.(\ref{eq.S.ex.Solid}):

\begin{equation}
S_{m,app}(N,M)=3Nln\left[ 1+\frac{M}{3N}\right] +Mln\left[ \frac{3N}{M}+1\right] .
\end{equation}

As in the two-level system, the approximate microcanonical and the canonical entropies expressed in terms of $M$ and $N$ come out the same \cite{Callen}: $S_{m,app}=S_{can}$. 

In Figure 5a, the three expressions for the entropy have been plotted for $N=10 $, i.e., $30$ oscillators. The discrepancy between the exact microcanonical entropy and the canonical entropy increases with energy. However, the relative error diminishes, see Table 2, from which it is clear that for $N=10$ and $M/3N \leq 10$ the error is greater than $5\%$. For
larger systems, the error can be neglected.

An explicit evaluation of $S_{m,ex}^{sur}$ and $S_{m,ex}^{vol}$ shows that the relative error between them is less than $2\%$ even for $N=10$ and $M/3N \leq 2$.

\subsection{Temperature}

To evaluate the exact temperature in the microcanonical formalism, the expression (\ref{eq.S.ex.Solid}) has to be used in (\ref{eq.temperature.finite}):

\begin{eqnarray}
T_{m,ex}(N,M)& =\left[ ln\left( \frac{(3N-1+M)!}{(3N-1)!M!}\right) -ln\left( \frac{(3N-2+M)!}{(3N-1)!(M-1)!}\right) \right]^{-1}\nonumber \\
&=\frac{1}{ln\left[ \frac{M+3N-1}{M}\right]} .
\label{eq.Tex.Solid}
\end{eqnarray}

The canonical temperature should be rewritten in terms of \textit{M}, i.e., \textit{E}, and \textit{N}, which are the ordinary variables of the microcanonical formalism. The canonical average energy $\bar{E}$ per oscillator is \cite{Kubo}

\begin{equation}
\bar{E}=\left( \frac{1}{e^{1/T}-1}\right).
\end{equation}

The average energy in microcanonical variables is $M/3N$. From the above equation, one has

\begin{equation}
T_{can}=\frac{1}{ln\left[ \frac{M+3N}{M}\right]}  .
\label{eq.T.can.Solido}
\end{equation}

Again, the exact microcanonical temperature differs from the canonical one by a term in $1/M$. Both temperatures are shown in Figure 5b for $N=10$: the differences between them are small. In Table 2, the relative error of the temperature is shown for different values of \textit{N} and \textit{M}: it slowly grows with the energy but diminishes strongly with increasing \textit{N}. Anyway, the calculated errors are less than $4\%$.

\subsection{Specific heat}

The specific heat in the microcanonical formalism is evaluated using Eqs. (\ref{eq.Cv.finite}) and (\ref{eq.Tex.Solid}). The expression for the canonical specific heat is \cite{Callen,Kubo,Reif}

\begin{equation}
C_{can}=3N\frac{1}{T^{2}}\frac{e^{1/T}}{\left( e^{1/T}-1\right) ^{2}}.
\end{equation}

The temperature has to be expressed in terms of the microcanonical variables according to (\ref{eq.T.can.Solido}) and replaced in the above equation; the result is

\begin{equation}
C_{can}=M\left( 1+\frac{M}{3N}\right) \left( ln\left[ 1+\frac{3N}{M}\right] \right) ^{2}.
\end{equation}

In Figure 5c, the canonical and exact microcanonical specific heats are plotted for $N=10$: both curves are almost identical. The relative error is relevant (around $5\%$) for $N=10$ and low energies as can be seen in Table 2, but it quickly diminishes with \textit{N} as expected.

Finally, in Figure 6, the relative error is plotted against the system size for a fixed energy $M/3N=2$. The error is lower than $4\%$ over the entire range.

\subsection{Summary}

The previous results can be summarized.

\begin{enumerate}
\item For $N=10$, the relative error between the exact surface microcanonical entropy and the approximate entropy is relevant for low energies (i.e., on the order of $3\%$) but it diminishes with increasing energy (for $M/3N=2$ it is $\sim 6\%$). The difference diminishes with larger system sizes. 
It should be remarked that the approximate entropy agrees with the canonical one.
\item The discrepancy between the volume and surface entropies is much lower than the error between the canonical entropy and the exact microcanonical entropy (on the order of 1\%--2\% over the plotted energy range).
\item The microcanonical and canonical temperatures are different for a small system ($N=10$) and high energies ($M/3N \geq 1.5$). The discrepancy between these temperatures increases with the energy (e.g., for $M/3N=0.1$, the relative error is $3\%$  and for $M/3N=2$, it is $\sim 3\%$).
\item The microcanonical and canonical specific heats differ for a small system ($N=10$) and low energy (for $M/3N=0.1$ it is $\sim 10\%$) but the difference decreases for larger systems.
\end{enumerate}

\section{Conclusions}
The aim of this paper has been to check some of  the usual assertions regarding the microcanonical formalism. Two well known models, the two-level system and the Einstein solid, have been solved exactly and those statements have been tested. The points taken into account are the following: the application of Stirling's approximation to deal with the logarithm of factorials, the use of derivatives instead of finite differences despite the discrete quantization of energy; the equivalence between the surface and volume entropy, and the equivalence between the results of the microcanonical and canonical formalisms.

The results for each model were summarized at the ends of Sections 3 and 4. The answers to the questions posed in Section 2 are

\textbf{Question 1}: For the two-level system, the volume entropy and the surface one differ for very small systems ($N = 10$) and energies $M/N \geq 0.2$ (Table 1). For the Einstein model, the difference between the two entropies is small even for a system with only a few atoms ($N = 10$).

\textbf{Question 2}: For the two-level system with few particles ($N = 10$), the exact microcanonical entropy and the canonical one (or the approximate microcanonical) do not agree. The relative error between them can be as high at $31\%$. For larger systems, the error decreases quickly.
For the Einstein solid, the exact microcanonical entropy and the canonical entropy are different for very small systems ($N = 10$) and low energies ($M/3N = 0.2$) and the relative error reaches values around $13\%$.
 
\textbf{Question 3}: For the two-level system, there is a relevant difference between the exact microcanonical temperature and specific heat and the canonical ones (Table 1) for small systems ($N = 10$). The conventional wisdom regarding the equivalence of both formalisms is not true for systems with a few tens of particles. Furthermore, there is a qualitative difference: the microcanonical temperature is well defined over the entire energy range ($0 \leq M/N < 0.5$), but the canonical temperature diverges at $M/N = 0.5$.
For the Einstein solid, the difference is less significant for small systems ($N = 10$) and negligible for systems with a few hundred atoms.
The discrepancies are larger in the two-level system than in the Einstein model because, for a solid with $N$ atoms, the number of oscillators  is $3N$ while in the first  model, $N$ particles means exactly $N$ two-level systems.

The final conclusion of this paper is that the assumptions mentioned in Section 2 are valid for systems with hundreds of particles. However, for systems with a few tens of particles, those statements lead to inexact results. The quantitative differences between the exact microcanonical and the canonical approaches can be as large as $45\%$ for very small systems ($N\sim 10$). Moreover, there are two qualitative differences for the two-level systems: the exact microcanonical approach shows that the temperature is well defined (no divergences) over the entire energy range and that the Schottky bump is shifted. Due to the ubiquity of the two-level system in physics, including paramagnetism in a solid\cite{Reif} and disordered systems\cite{Ziman}, these results might be of interest to the research community dealing with systems of a few tens of particles. 

Although these results refer to particular models, they are a step towards a general answer to the questions posed. Moreover, the methodology used is applicable to other well-known models that are exactly solvable in the microcanonical formalism. Work is in progress in this direction. 

\section*{Acknowledgments}
The authors thank the National Scientific and Technological Research Council (CONICET) of Argentina for financial support. DSB also acknowledge partial support from project SeCTyP 06/M035.

\newpage   

\begin{table}[h!]
\centering
\caption{Two-state system: relative errors in percent between the canonical and microcanonical results for different system sizes \textit{N} and energies \textit{M}. The first column refers to the entropy, the second to the temperature, and the third to the specific heat.}
\begin{indented}
\item[]\begin{tabular}{|c c|c c c|c c c|c c c|}
\cline{3-11} 
\multicolumn{1}{c}{} &  & \multicolumn{9}{c|}{\textbf{N}}\tabularnewline
\cline{3-11} 
\multicolumn{1}{c}{} &  & \multicolumn{3}{c|}{\textbf{S}} & \multicolumn{3}{c|}{\textbf{T}} & \multicolumn{3}{c|}{\textbf{C}}\tabularnewline
\multicolumn{1}{c}{} &  & \textit{10} & \textit{100} & \textit{200} & \textit{10} & \textit{100} & \textit{200} & \textit{10} & \textit{100} & \textit{200}\tabularnewline
\hline 
\multirow{3}{*}{\textbf{M/N}} & \textit{0.2} & 31.5 &  4.84& 2.72 & 8.5 & 0.9 & 0.4 &9.06  & 0.61 &0.30 \tabularnewline
 &\textit{0.3} & 27.6 & 4.17 & 2.34 & 15.8 & 1.7 & 0.8 & 15.7 &0.93  &0.45 \tabularnewline
 & \textit{0.4} &25.9  & 3.87 & 2.17 & 38 & 4.1 & 2.0 & 45.9 & 1.22 &0.55 \tabularnewline
\hline 
\end{tabular}
\end{indented}
\end{table}

\begin{table}[h!]
\centering
\caption{Einstein model of a solid: relative errors in percent between the canonical and microcanonical results for different system sizes \textit{N} and energies \textit{M}. The first column refers to the entropy, the second to the temperature, and the third to the specific heat.}
\begin{indented}
\item[]\begin{tabular}{|c c|c c c |c c c|c c c|}
\cline{3-11} 
\multicolumn{1}{c}{} &  & \multicolumn{9}{c|}{\textbf{N}}\tabularnewline
\cline{3-11} 
\multicolumn{1}{c}{} &  & \multicolumn{3}{c|}{\textbf{S}} & \multicolumn{3}{c|}{\textbf{T}} & \multicolumn{3}{c|}{\textbf{C}}\tabularnewline
\multicolumn{1}{c}{} &  & \textit{10} & \textit{100} & \textit{200} & \textit{10} & \textit{100} & \textit{200} & \textit{10} & \textit{100} & \textit{200}\tabularnewline
\hline 
\multirow{3}{*}{\textbf{M/3N}} & \textit{0.2} & 13.4 &  1.9& 1.06 & -1.6 & -0.15 & -0.08 &4.8  & 0.45 &0.23 \tabularnewline
 &\textit{1} & 7.7 & 1.0 & 0.54 & -2.4 & -0.24 & -0.12 & 3.3 &0.32  &0.16 \tabularnewline
 & \textit{30} &5.2  & 0.6 & 0.03 & -3.2 & -0.32 & -0.14 & 3.4 & 0.33 &0.17 \tabularnewline
\hline 
\end{tabular}
\end{indented}
\end{table}

\newpage   

\section*{Figures and Figure captions}

\begin{figure}[h!]
\centering
\includegraphics[width=0.7\textwidth]{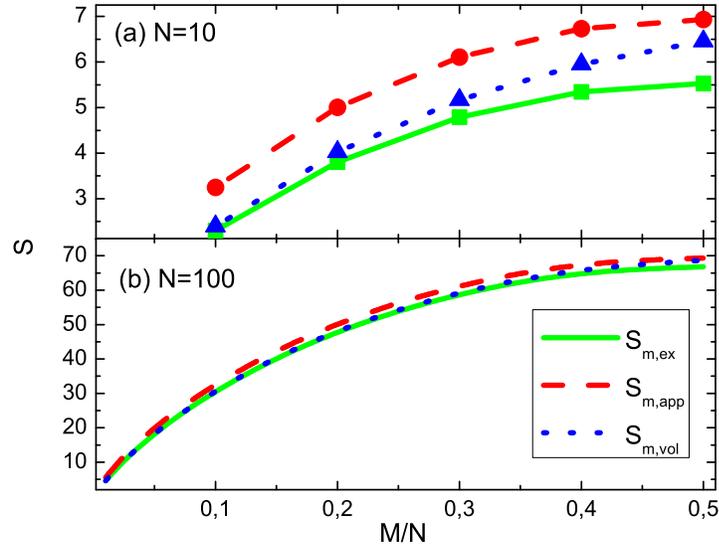}   
\caption{Entropy of a two-level system in terms of the average energy for two different system sizes (a) $N=10$ and (b) $N=100$. The continuous line is the surface entropy evaluated in the microcanonical formalism $S_{m}^{sur}$ without any approximation. The dotted line is the exact volume entropy $S_{m}^{vol}$ in the microcanonical framework. The dashed line is the approximate entropy in the microcanonical formalism $S_{m,app}$ that is found using Stirling's formula. It can be shown that $S_{m,app} = S_{can}$ where $S_{can}$ is the canonical entropy. For smaller systems, the differences between the three entropies are relevant. It cannot be said that the canonical and microcanonical approaches lead to the same result.}
\label{}
\end{figure}

\begin{figure}[h!]
\centering
\includegraphics[width=0.7\textwidth]{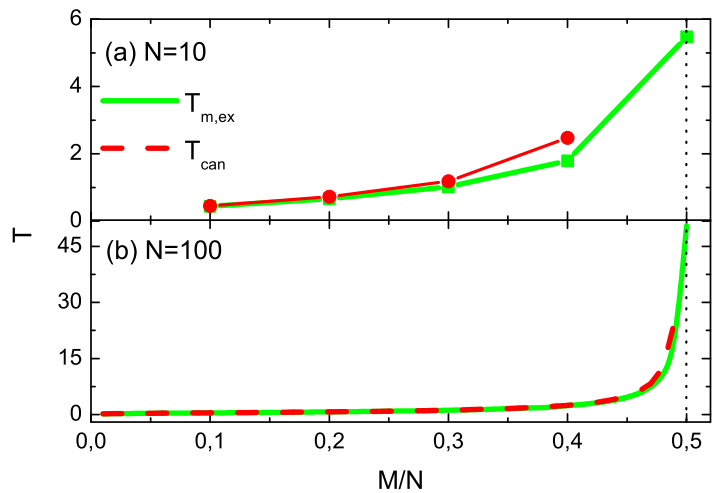}   
\caption{Temperature of the two-level system as a function of the average energy for (a) $N=10$ elements and (b) $N=100$ elements. The continuous line is the exact temperature evaluated in the microcanonical formalism, i.e.,
Stirling's approximation was not used and the derivative used to get \textit{T} from \textit{S} was replaced by finite differences because of the discrete nature of the energy. The dashed line is the temperature obtained with the canonical formalism. There are large discrepancies between the two curves for $M/N>0.4$. For $M/N=0.5$, the canonical temperature diverges while the microcanonical one has a well defined value.}
\label{}
\end{figure}

\begin{figure}[h!]
\centering
\includegraphics[width=0.7\textwidth]{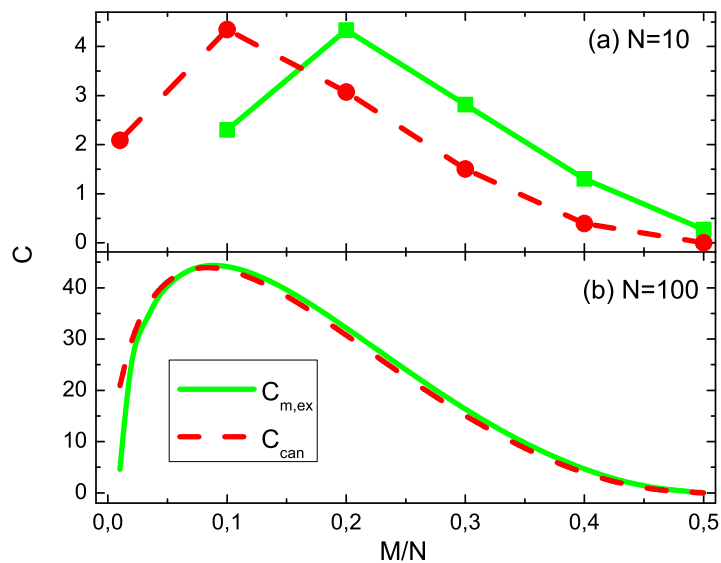}   
\caption{Specific heat for a two-level system in terms of the average energy for (a) $N=10$ and (b) $N=100$. The continuous line is the exact specific heat evaluated in the microcanonical formalism; the derivative needed to get \textit{C} has been replaced by a finite difference due to the quantization of energy. The dashed line is the specific heat evaluated in the canonical ensemble. For the smaller size, there is an appreciable difference between the two curves and a shift in the position of the Schottky bump.}
\label{}
\end{figure}

\begin{figure}[h!]
\centering
\includegraphics[width=0.7\textwidth]{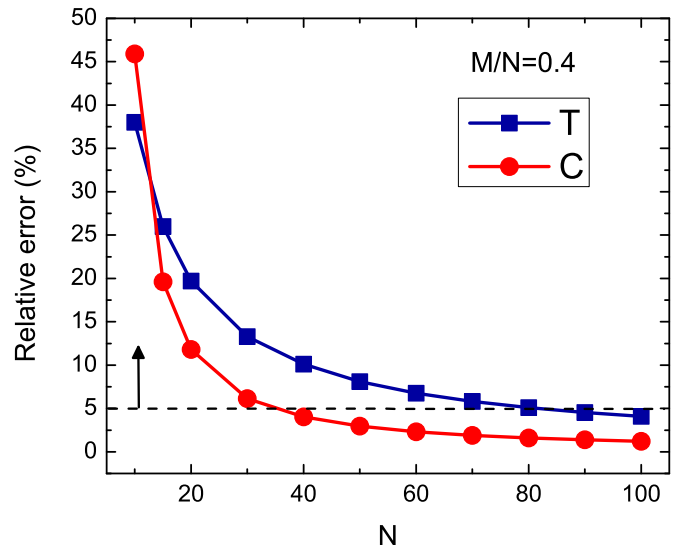}   
\caption{Relative errors of the temperature and the specific heat evaluated in the canonical and microcanonical frameworks. They are plotted against the system size for a fixed average energy $M/N=0.4$. The dashed line corresponds to a $5\%$ error. For systems larger than $N=200$, the two formalisms give close results.}
\label{}
\end{figure}

\begin{figure}[h!]
\centering
\includegraphics[width=0.7\textwidth]{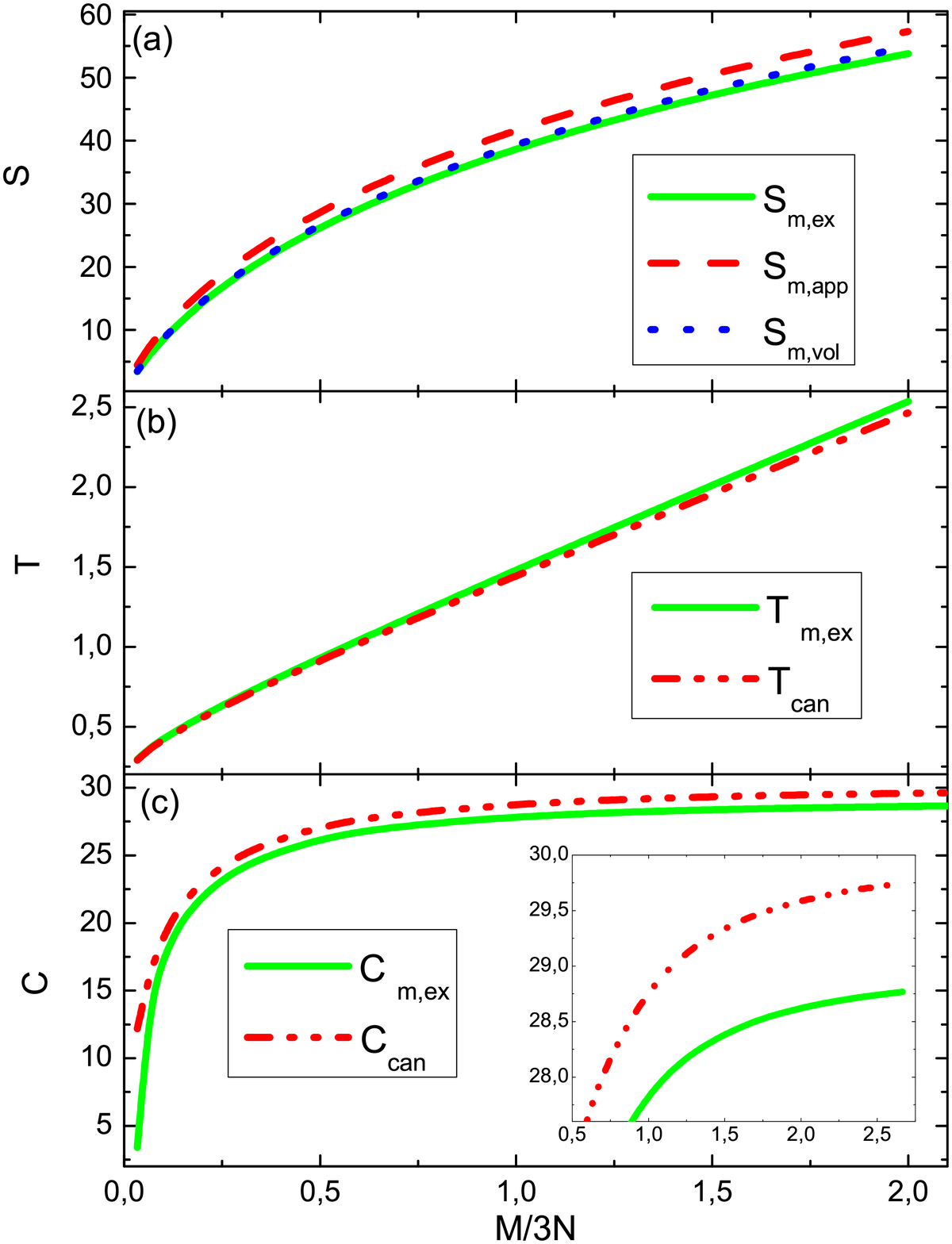}   
\caption{Thermodynamic variables of the Einstein model in terms of the average energy per oscillator for a system size of $N=10$ atoms or 30 oscillators. 
(a) \textbf{Entropy}. The solid (dotted) line is the surface (volume) entropy evaluated in the microcanonical formalism without any approximation. The dashed line is the approximate microcanonical entropy, which comes out equal to the entropy evaluated in the canonical formalism. The discrepancy between the canonical and microcanonical entropies is noticeable for $M/3N > 1$
(b) \textbf{Temperature}.  The solid line is the microcanonical temperature while the dashed line is the canonical temperature expressed in terms of the energy. The discrepancy between these temperatures is lower than $\sim 3\%$.
(c) \textbf{Specific heat}. The solid (dashed) lined is the value obtained with the microcanonical (canonical) formalism. The relative error is between $3\%$ and $5\%$ }
\label{}
\end{figure}

\begin{figure}[h!]
\centering
\includegraphics[width=0.7\textwidth]{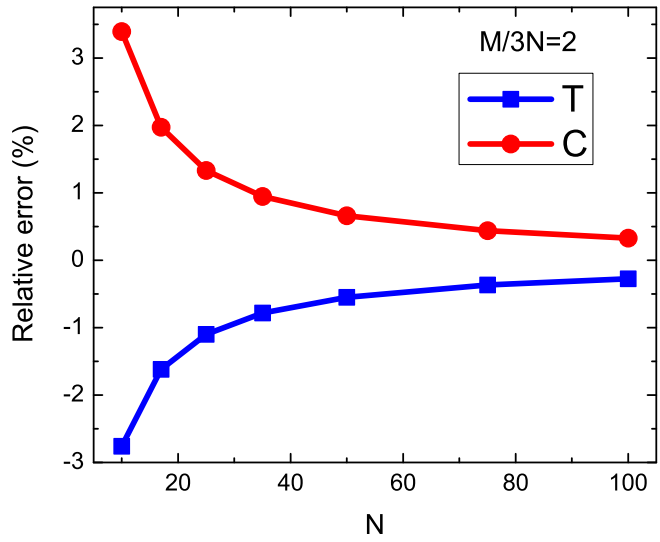}   
\caption{Einstein model. The relative errors between the canonical and microcanonical results as a function of the system size. The average oscillator energy is fixed at $M/3N=2$. The errors are lower than $4\%$ even for very small systems.}
\label{}
\end{figure}

\end{document}